\documentclass[showpacs,aps,twocolumn]{revtex4}
\usepackage{amsmath,bm}
\usepackage{epsfig}
\usepackage{graphicx}
\usepackage{amssymb}
\usepackage{lscape}
\usepackage{booktabs}
\usepackage{subfig}

\begin{document}
\title{Systematic study for particle transverse momentum asymmetry \\ in
       minimum bias pp collisions at LHC energies}
\author{Yu-Liang Yan$^{1,2,3}$ \footnote{yanyl@ciae.ac.cn}, Dai-Mei Zhou$^4$,
        Ayut Limphirat$^{1,3}$, Bao-Guo Dong$^2$, Yu-Peng Yan$^{1,3}$, and
        Ben-Hao Sa$^{1,2,4}$ \footnote{sabh@ciae.ac.cn}}

\affiliation{$^1$ School of Physics, Institute of Science, Suranaree
              University of Technology, Nakhon Ratchasima 30000, Thailand.\\
             $^2$ China Institute of Atomic Energy, P. O. Box 275 (10),
              Beijing, 102413 China.\\
             $^3$ Thailand Center of Excellence in Physics (ThEP), Commission
              on Higher Education, Bangkok 10400, Thailand.\\
             $^4$ Institute of Particle Physics, Central China Normal
              University, 430082 Wuhan, China \\ and Key Laboratory of Quark
              and Lepton Physics (CCNU), Ministry of Education, China.}

\begin{abstract}
A method of randomly rearranging the $p_x$ and $p_y$ components of the
produced particle on the circumference of an ellipse with the half major and
minor axes being $p_T(1+\delta_p)$ and $p_T(1-\delta_p)$ is introduced. The
ALICE data on the transverse sphericity as a function of charged multiplicity
in the minimum bias pp collisions at $\sqrt s$=0.9 and 7 TeV are well
reproduced by this method based on the particles generated in the PYTHIA6.4
simulations. The correspondingly predicted charged particle $v_2$ upper limit
is a measurable value of $\sim$0.2-0.3 . We suggest a systematic measurements
for the particle transverse momentum sphericity and the elliptic flow
parameter.
\end{abstract}
\date{\today}
\maketitle
\section{Introduction}
The event shapes are relevant to the properties of hadronic final state.
One has employed the hadronic event shape measurements to test asymptotic
freedom and to extract the strong coupling constant, etc. in the $e^+e^-$
annihilation and lepton deep inelastic scattering, respectively, for a long
time \cite{opal,zeus}. Recently, the final hadronic event shapes in pp
collisions at the LHC energies have been measured by CMS \cite{cms}, ALICE
\cite{alice1,alice}, and ATLAS \cite{atlas,atlas1}.

In order to avoid bias from the boost along beam axis \cite{banfi}, this study
is restricted to the transverse momentum plane. We start from the transverse
momentum matrix of the produced charged particles \cite{alice}
\begin{equation*}
{\bf S_{xy}}=\left(  \begin{array}{cc}
a_{11} & a_{12} \\
a_{21} & a_{22}
\end{array}  \right),
\end{equation*}
\begin{equation*}
a_{11}=\frac{1}{\sum_i p_{T_i}}\sum_i\frac{p_{x_i}^2}{p_{T_i}},
\end{equation*}
\begin{equation*}
a_{22}=\frac{1}{\sum_i p_{T_i}}\sum_i\frac{p_{y_i}^2}{p_{T_i}},
\end{equation*}
\begin{equation}
a_{12}=a_{21}=\frac{1}{\sum_i p_{T_i}}\sum_i\frac{p_{x_i}p_{y_i}}{p_{T_i}},
\label{equ1}
\end{equation}
where $p_{T_i}$ is the transverse momentum of particle $i$, $p_{x_i}$ and
$p_{y_i}$ are the corresponding transverse momentum components, and
the sum runs over the charged particles in a single event. The two eigenvalues of
this transverse momentum matrix
%\begin{equation}
%\lambda_{1,2}=\frac{(a_{11}+a_{22})\pm[(a_{11}-a_{22})^2+4a_{12}^2]^{1/2}}{2}.
%\end{equation}
satisfy
\begin{equation}
\lambda_1+\lambda_2=a_{11}+a_{22}=1.
\end{equation}
If they are ordered in $\lambda_1>\lambda_2$, the transverse momentum
sphericity is then defined as \cite{alice}
\begin{equation}
S_T= 2\lambda_2 .
\label{spher}
\end{equation}
By intuitive construction one knows that $S_T$ possesses limits of
\begin{equation}
S_T=\left\{\begin{array}{lll}
        0, &  {\rm pencil-like\;  limit,} \\
        1, & {\rm isotropic\;  limit,}
       \end{array}
 \right.
\label{limi1}
\end{equation}
and hence $S_T$ is a measure of the degree of transverse momentum
azimuthal symmetry. The event averaged transverse momentum sphericity is then
denoted as $\langle S_T\rangle$, where $\langle ...\rangle$ indicates an
average over events.

As mentioned in \cite{olli} that the transverse momentum azimuthal asymmetry
is measured by the dimensionless observable
\begin{equation}
A_T=\frac{\lambda_1-\lambda_2}{\lambda_1+\lambda_2}\\
 =[(a_{11}-a_{22})^2+4a_{12}^2]^{1/2}.
\label{assy}
\end{equation}
It is easy to prove
\begin{equation}
S_T+A_T=1.
\label{limi3}
\end{equation}
Corresponding to the $S_T$ limits in Eq.~(\ref{limi1}) $A_T$ possesses
the limits of
\begin{equation}
A_T=\left\{\begin{array}{lll}
        1, &  {\rm pencil-like\; limit,} \\
        0, & {\rm isotropic\; limit,}
       \end{array}
\label{limi2}
\right.
\end{equation}
and therefore $A_T$ is also a measure of the degree of transverse momentum azimuthal
asymmetry. The event averaged transverse momentum asymmetry is denoted as
$\langle A_T\rangle$.
\begin{figure*}[htbp]
\centering
\includegraphics[width=0.75\textwidth,clip=true]{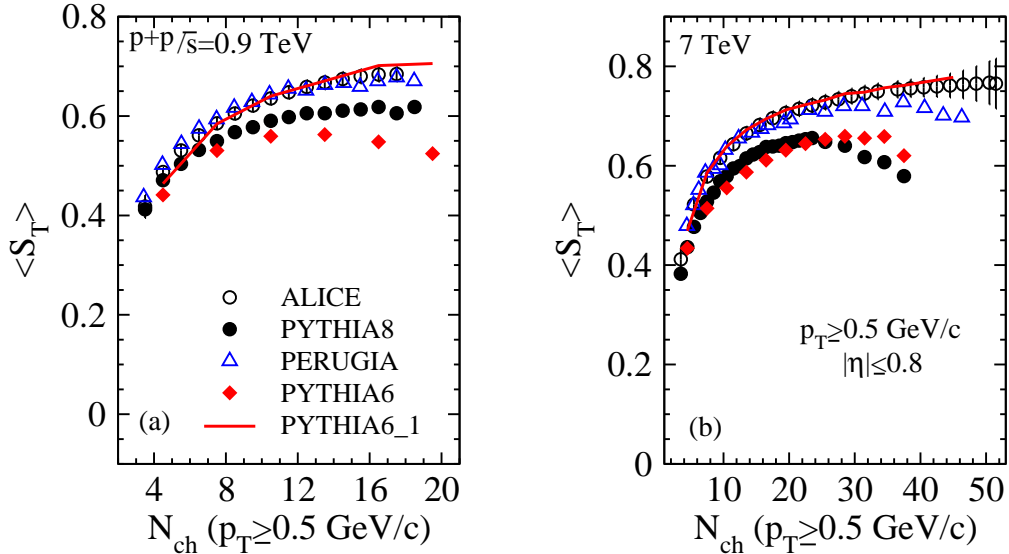}
\caption{(Color online) Charged particle event averaged transverse
         sphericity as a function of charged multiplicity in the
         pp collisions at (a) $\sqrt s$=0.9 TeV and (b) $\sqrt s$=7 TeV. The
         ALICE data are taken from \cite{alice} and the PYTHIA8 as well as
         PERUGIA results are copied from \cite{alice}.}
\label{spheri_2}
\end{figure*}
\section{Models and simulations}
\subsection{Sphericity from PYTHIA6.4}
PYTHIA6.4 \cite{sjo} is used to calculate the charged particle
transverse momentum sphericity as a function of charge multiplicity,
$\langle S_T\rangle(N_{ch})$, in $p_T\geq$0.5 GeV/c and $|\eta|\leq$0.8, for
the minimum bias pp collisions at $\sqrt s$=0.9 and 7 Tev. These results
(red solid diamonds, indicated as PYTHIA6) are compared with the corresponding
ALICE data of ``all" events (black open circles, indicated as ALICE)
\cite{alice} in Fig.~\ref{spheri_2} (a) and (b) for 0.9 and 7 TeV,
respectively. The black solid circles indicated as PYTHIA8 in this figure are
the results of PYTHIA8 copied from \cite{alice}. One sees here that the
results of PYTHIA6, like PYTHIA8, are not consistent with the ALICE data.

One way out is to invoke the various PYTHIA6 tunes. They are based on
PYTHIA6.4 but with a couple of extra model parameters, relevant to the soft-
and/or hard-processes as well as the parton distribution function etc.,
fitting to the experimental data. For example, in the PERUGIA-2011 tune
\cite{prd} ``the data sets used to constrain the models include hadronic
$Z^0$ decay at LEP, Tavatron min-bias data at 630, 1800, and 1960 GeV,
Tevatron Drell-Yan data at 1800 and 1960 GeV, and SPS min-bias data at 200,
546, and 900 GeV". As mentioned in \cite{alice}, the above
ALICE data were best reproduced by the PERUGIA-2011 tune (cf. the blue open
triangles indicated as PERUGIA in Fig.~\ref{spheri_2}) among the PYTHIA6
tunes.

\subsection{Rearrangement method}
We prefer to have a more simplified modification for the PYTHIA6.4 model. A
rearrangement method is proposed, in which the $p_x$ and $p_y$
components of produced particle from the PYTHIA6.4 simulations are randomly
rearranged on the circumference of an ellipse with the half major and minor
axes of
\begin{equation}
a=p_T(1+\delta_p) \quad b=p_T(1-\delta_p),
\label{axi}
\end{equation}
where $p_T$ is the transverse momentum of the produced particle and
$0<\delta_p<1$ is an extra introduced deformation parameter. This is
equivalent to randomly re-sample $p_x$ and $p_y$ of the produced particle
according to
\begin{equation}
p_x=p_T(1+\delta_p)\cos\phi, \quad p_y=p_T(1-\delta_p)\sin\phi,
\label{axi1}
\end{equation}
as the equation of ellipse
\begin{equation}
\frac{p_x^2}{a^2}+\frac{p_y^2}{b^2}=1
\end{equation}
is held, where the half major and minor axes are given by Eq.~(\ref{axi}). In the Eq.~
(\ref{axi1}) $\phi$ is randomly distributed in [0, 2$\pi$]. Then we may fit the deformation parameter $\delta_p$ to the (ALICE) sphericity data and make
prediction for other observables such as the elliptic flow parameter
$v_2$, or vice versa.

As the particle transverse momentum may be change in the rearrangement, one
has to re-calculate the transverse momentum by $p_T^2=p_x^2+p_y^2$ after the
rearrangement. However we will see later that provided $\delta_p$ is far
less than unity (small perturbation) the change in particle transverse
momentum distribution caused by the rearrangement is not visual.
\begin{figure*}[htbp]
\centering
\includegraphics[width=0.7\textwidth,clip=true]{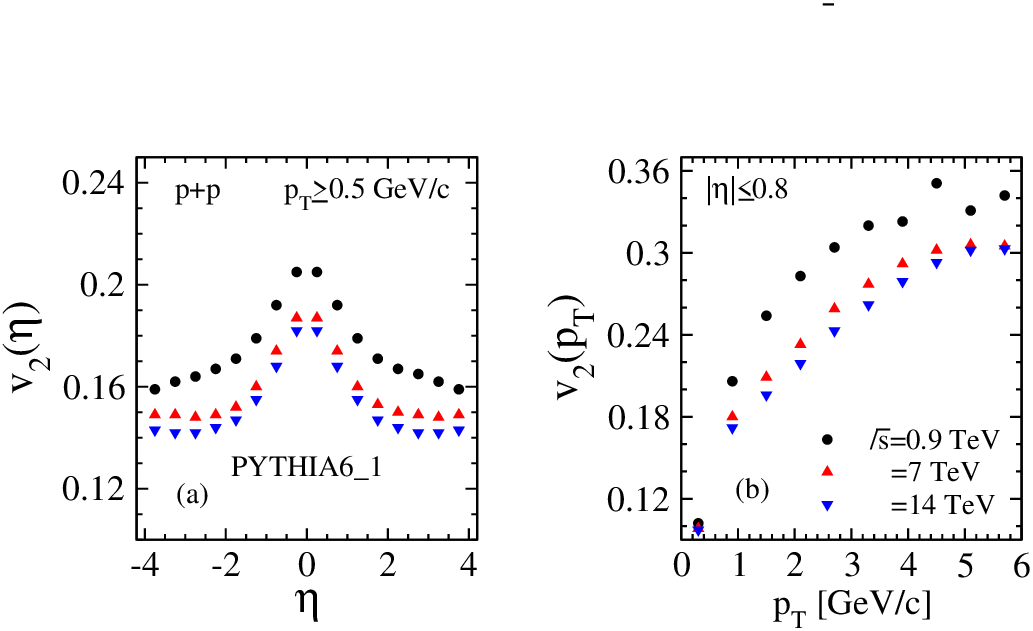}
\caption{(Color online) Charged particle $v_2(\eta)$ (a) and $v_2(p_T)$ (b)
          in the minimum bias pp collisions at LHC energies resulted from
          PYTHIA6\_1 simulations.}
\label{v2etapt_9007pp}
\end{figure*}
\subsection{Fitting process}
The fitting process could be started from the small $\delta_p$ side,
$\delta_p=0.01$ for instance. If the calculated $\langle S_T\rangle(N_{ch})$
is larger than the ALICE data we increase the $\delta_p$ value and repeat the
calculation, vice versa otherwise. At last, we obtain the results (red
curves indicated PYTHIA6\_1 in Fig.~\ref{spheri_2}) calculated with
$\delta_p$=0.092 and 0.091, consistent well with the ALICE data of
$\langle S_T\rangle(N_{ch})$ in the minimum bias pp collisions at $\sqrt{s}$=
0.9 and 7 TeV, respectively.

\section{Predictions for the elliptic flow parameter}
The Fourier expansion is another method investigating the particle transverse
momentum azimuthal asymmetry \cite{zhang,posk}. In \cite{posk} the Fourier
expansion of the particle number (multiplicity) distribution is expressed as
\begin{equation}
E{\frac{d^3N}{d^3p}}=\frac{1}{2\pi}\frac{d^2N}{p_Tdydp_T}
       [1+\sum_{n=1,2,...}2v_ncos[n(\phi-\Psi_r)]],
\label{f1}
\end{equation}
where $\phi$ refers to the azimuthal angle of particle transverse momentum,
$\Psi_r$ stands for the azimuthal angle of reaction plane. In the theoretical
study, if the beam direction and impact parameter vector are fixed,
respectively, on the $p_z$ and $p_x$ axes in the laboratory (Lab.) frame, then
the reaction plane is just the $p_x-p_z$ plane \cite{zhang}. Therefore the
reaction plane angle, $\Psi_r$, between the reaction plane and $p_x$ axis
\cite{zhang} introduced for the extracting elliptic flow experimentally
\cite{posk} is zero. The equation (\ref{f1}) and the harmonic coefficients
there reduce to
\begin{align}
E{\frac{d^3N}{d^3p}}=&\frac{1}{2\pi}\frac{d^2N}{p_Tdydp_T}
       [1+\sum_{n=1,2,...}2v_ncos(n\phi)], \nonumber\\
\langle v_n\rangle_p=&\langle cos(n\phi)\rangle_p, \nonumber\\
\langle v_1\rangle_p=&\langle\frac{p_x}{p_T}\rangle_p, \nonumber\\
\langle v_2\rangle_p=&\langle\frac{p_x^2-p_y^2}{p_T^2}\rangle_p, \nonumber\\
 . . .
\label{f2}
\end{align}
where $\langle ... \rangle_p$ denotes the particle-wise average, i.e. averaged
over all particles in all events \cite{posk}.
\begin{figure*}[htbp]
\centering
\includegraphics[width=0.65\textwidth ,clip=true]{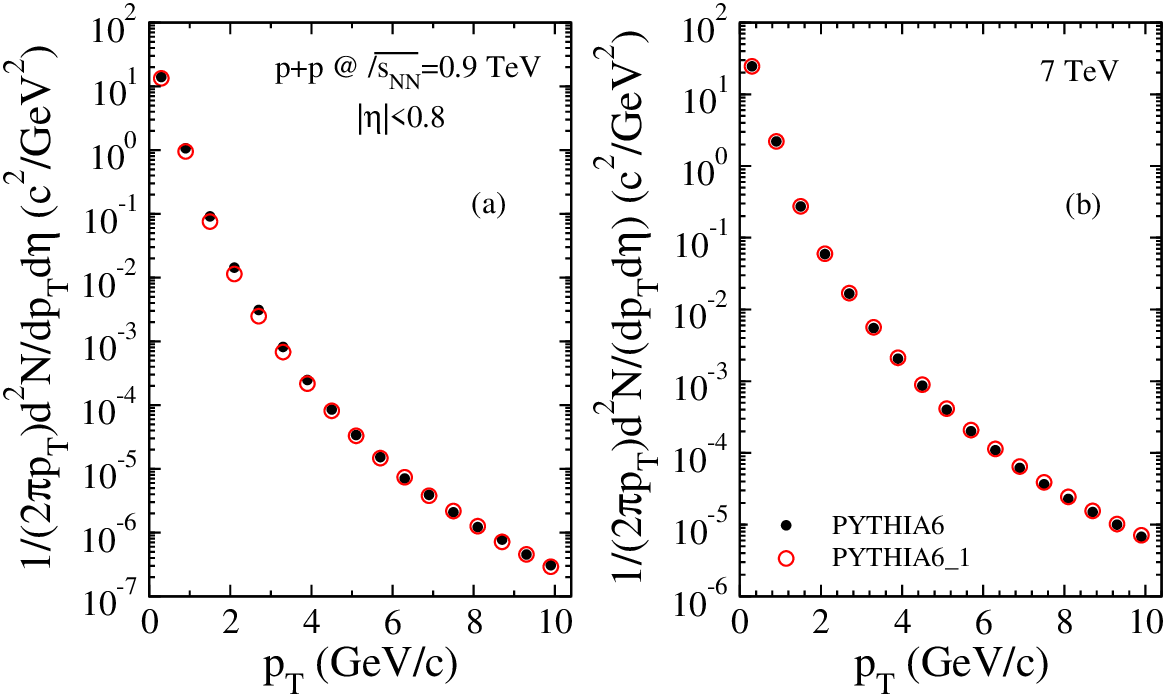}
\caption{(Color online) Charged particle transverse momentum distributions
          in the minimum bias pp collisions at $\sqrt s$=0.9 (a) and 7 TeV
          (b).}
\label{pt9007}
\end{figure*}

The second harmonic coefficient (elliptic flow parameter $v_2$) is specially
important because the large $v_2$ of emitted particles is a characteristic
feature of the hot and dense medium created in the ultra-relativistic nuclear
collisions. It has contributed to the suggestion of a strongly coupled
quark-gluon plasma (sQGP) observed in the nucleus-nucleus collisions at RHIC
energies \cite{brah,phob1,star,phen}.

There are a number of predictions for the elliptic flow parameter $v_2$ in
high energy (multiplicity) pp collisions \cite{wied,chau,boze,zhou}. We pursue
having a prediction based on the experiments, the sphericity measurements
\cite{cms,alice1,alice,atlas}.
\begin{table*}[htbp]
\caption{$v_2$ upper limit obtained by averaging the $v_2(p_T)$ over $p_T$ in
         the minimum bias pp collisions at $\sqrt s$=0.9, 7 and 14 TeV.}
\centering
\begin{tabular}{cccccccc}
\hline\hline
\cmidrule[0.25pt](l{0.05cm}r{0.05cm}){1-1}
\cmidrule[0.25pt](l{0.05cm}r{0.05cm}){2-3}
\cmidrule[0.25pt](l{0.55cm}r{0.55cm}){4-4}
\cmidrule[0.25pt](l{0.05cm}r{0.05cm}){5-6}
\cmidrule[0.25pt](l{0.05cm}r{0.05cm}){7-7}
\cmidrule[0.25pt](l{0.05cm}r{0.05cm}){8-8}
&\multicolumn{2}{c}{$\sqrt s$=0.9 TeV}& &\multicolumn{2}{c}{$\sqrt s$=7 TeV}
& &$\sqrt s$=14 TeV\\
\hline
& PYTHIA6& PYTHIA6\_1& & PYTHIA6& PYTHIA6\_1& & PYTHIA6\_1\\
partial$^\dag$& $\sim$0& 0.297& & $\sim$0& 0.275& &0.267  \\
full$^\ddag$  & $\sim$0& 0.291& & $\sim$0& 0.243& &0.237  \\
\hline\hline
\multicolumn{8}{l}{$^\dag$ calculated in partial phase space of $p_T\geq$0.5
GeV/c and $|\eta|\leq0.8$}.\\
\multicolumn{8}{l}{$^\ddag$ calculated in full $p_T$ and $\eta$ phase space
                  }.\\
\end{tabular}
\label{sphecc}
\end{table*}

The predictions of the above
PYTHIA6\_1 simulations are given in Fig. \ref{v2etapt_9007pp} (a) and (b) for
the
charged particle $v_2(\eta)$ and $v_2(p_T)$ in the minimum bias pp collisions
at $\sqrt s$=0.9,
7 and 14 TeV (which is calculated with $\delta_p$=0.091 in addition),
respectively. Table~\ref{sphecc} gives the upper limit of the charged particle
elliptic flow parameter $v_2$ obtained by averaging $v_2(p_T)$ over $p_T$.
It is found in the table that the $v_2$ upper limits are estimated as 0.2-0.3 for
minimum bias pp collisions at the LHC energies, which is consistent
with the existing results of 0.04 - 0.2 in \cite{wied,chau,boze,zhou}. We also see
in the table that the $v_2$ upper limit decreases with increasing
the reaction energy. This is consistent
with the ALICE results that the charged particle event averaged transverse
sphericity increases from 0.613 at 0.9 TeV to 0.700 at 7 TeV.

To check how the
momentum rearrangement defined in  Eq.~(\ref{axi}) change the transverse momentum
distribution, we calculate the charged particle transverse momentum
distributions. Shown in Fig.~\ref{pt9007} are the results of
PYTHIA6 (black solid circles) and PYTHIA6\_1 (red open
circles) simulations for the minimum bias pp collisions at $\sqrt s$=0.9 (a) and 7 TeV
(b). It is found that
the results of PYTHIA6 and PYTHIA6\_1 are so close that one may conclude that provided
the deformation parameter $\delta_p$ is far less than unity (small
perturbation) the change in transverse momentum distribution caused by the
rearrangement is trivial.

\section{Discussion and conclusion}
In summary, we have proposed a method of randomly rearranging the $p_x$ and
$p_y$ components of produced particle from the PYTHIA6.4 simulation on the
circumference of an ellipse with the half major and minor axes being
$p_T(1+\delta_p)$ and $p_T(1-\delta_p)$. The charged particle transverse
momentum sphericity and the elliptic flow parameter in the minimum bias pp
collisions at $\sqrt{s}$=0.9 and 7 TeV are then calculated systematically.
The ALICE data of event averaged charged particle transverse sphericity as a
function of charged multiplicity in the above pp collisions are well
reproduced. The elliptic flow parameter as a function of $\eta$ ($v_2(\eta)$)
and $p_T$ ($v_2(p_T)$) as well as the $v_2$ upper limits of $\sim$ 0.2-0.3
obtained by averaging $v_2(p_T)$ over $p_T$ are predicted for the minimum
bias pp collisions at the LHC energies.

As mentioned in \cite{atlas} that the charged particle transverse momentum
sphericity is measured (defined) ``using jets to represent the final state
four-momentum." This measurement is only influenced by the jet reconstruction.
However, the measurement of $v_2$, whatever the event plane method \cite{posk}
or the Lee-Yang zero point method \cite{lyzp} or the cumulant method
\cite{cumu}, is much more model dependence, such as the dependence on the
model selected for the nonflow decomposition \cite{wang}. The cumulant
method is even distinguished by two-, four-, and six-particle cumulants. So
the discrepancy among the measured $v_2$ values with the different methods may
reach 10-100\% \cite{star1,cms1}. Recently, one even argued that the event
plane method is obsolete \cite{luzu}. Therefore, we strongly suggest that the
transverse momentum sphericity and the elliptic flow parameter should be
measured simultaneously for the benefit of cross checking and the reliable
measurements.

Acknowledgements: This work was supported by the National Natural Science
Foundation of China under grant Nos.:11075217, 11105227, 11175070,
11477130 and by the 111 project of the foreign expert bureau of China. AL
and YPY acknowledge the financial support from TRF-CHE-SUT under contract
No. MRG5480186. YLY acknowledges the financial support from SUT-NRU project
under contract No. 17/2555. BHS would like to thank Prof. T. Sj\"{o}strand
for the helps continuously.

\end{document}